\documentclass[12pt,preprint]{aastex}

\begin{document}

\newcommand{\lya}{Lyman-$\alpha$}          
\newcommand{\eqw}{\hbox{EW}}
\newcommand{\be}{\begin{equation}}     
\newcommand{\ee}{\end{equation}}
\def\Mmin{{\rm  M}_{\rm min}}  
\def\erg{\hbox{erg}} 
\def\cm{\hbox{cm}}
\def\sec{\hbox{s}}        
\def\f17{f_{17}}        
\def\Mpc{\hbox{Mpc}}
\def\cMpc{\hbox{cMpc}}     
\def\Gpc{\hbox{Gpc}}     
\def\nm{\hbox{nm}}
\def\km{\hbox{km}}  
\def\kms{\hbox{km  s$^{-1}$}} 
\def\year{\hbox{yr}}
\def\Myr{\hbox{Myr}}     
\def\Gyr{\hbox{Gyr}}     
\def\deg{\hbox{deg}}
\def\arcsec{\hbox{arcsec}}   
\def\microJy{\mu\hbox{Jy}}  
\def\zre{z_r}
\def\fesc{f_{\rm esc}}

\def\ergcm2s{\ifmmode                  {\rm\,erg\,cm^{-2}\,s^{-1}}\else
                ${\rm\,ergs\,cm^{-2}\,s^{-1}}$\fi}
                \def\ergsec{\ifmmode            {\rm\,erg\,s^{-1}}\else
                ${\rm\,ergs\,s^{-1}}$\fi}
                \def\kmsMpc{\ifmmode{\rm\,km\,s^{-1}\,Mpc^{-1}}\else
                ${\rm\,km\,s^{-1}\,Mpc^{-1}}$\fi}  \def\kpc{{\rm kpc}}
                \def\nv{\ion{N}{5}  $\lambda$1240} \def\civ{\ion{C}{4}
                $\lambda$1549}   \def\oii{[\ion{O}{2}]  $\lambda$3727}
                \def\oiipair{[\ion{O}{2}]  $\lambda \lambda$3726,3729}
                \def\oiii{[\ion{O}{3}]                   $\lambda$5007}
                \def\oiiib{[\ion{O}{3}]                  $\lambda$4959}
                \def\oiiipair{[\ion{O}{3}] $\lambda \lambda$4959,5007}
                \def\taulya{\tau_{Ly\alpha}}
                \def\taubar{\bar{\tau}_{Ly\alpha}}
                \def\llya{L_{Ly\alpha}}                \def\ldlya{{\cal
                L}_{Ly\alpha}}  \def\nbar{\bar{n}}  \def\Msun{M_\odot}
                \def\sqamin{\Box'}

\title{Clustering of Lyman alpha emitters at z $\approx$ 4.5}

\author{Katarina Kova\v{c}$^{1,2}$, Rachel S. Somerville$^3$, 
James E. Rhoads$^{4,5}$,
Sangeeta Malhotra$^{4,5}$, JunXian Wang$^6$}

\begin{abstract}
We present the clustering properties of 151 \lya\ emitting galaxies at
$z \approx 4.5$ selected from the Large Area Lyman Alpha (LALA)
survey. Our catalog covers an area of 36' x 36' observed with five
narrowband filters. We assume that the angular correlation function
$w(\theta)$ is well represented by a power law $A_{w} \Theta^{-\beta}$
with slope $\beta = 0.8$, and we find $A_w = 6.73 \pm 1.80$. We then
calculate the correlation length $r_0$ of the real-space two-point
correlation function $\xi(r) = (r/r_0)^{-1.8}$ from $A_w$ through the
Limber transformation, assuming a flat, $\Lambda$-dominated
universe. Neglecting contamination, we find $r_0 = 3.20 \pm 0.42$
$h^{-1}$ Mpc. Taking into account a possible 28\% contamination by
randomly distributed sources, we find $r_0=4.61 \pm 0.6$ $h^{-1}$
Mpc. We compare these results with the expectations for the clustering
of dark matter halos at this redshift in a Cold Dark Matter model, and
find that the measured clustering strength can be reproduced if these
objects reside in halos with a minimum mass of 1--$2\times 10^{11}
h^{-1} M_{\odot}$.  Our estimated correlation length implies a bias
of $b\sim3.7$, similar to that of Lyman-break galaxies (LBG) at
$z\sim3.8-4.9$. However, \lya\ emitters are a factor of $\sim 2$--16
rarer than LBGs with a similar bias value and implied host halo
mass. Therefore, one plausible scenario seems to be that \lya\ emitters
occupy host halos of roughly the same mass as LBGs, but shine with a
relatively low duty cycle of 6--50\%.
\end{abstract}

\altaffiltext{1}{Kapteyn   Astronomical   Institute,   University   of
Groningen,  P.O.Box 800,  9700 AV  Groningen, The  Netherlands}
\altaffiltext{2}{Present address: Department of Physics, Swiss Federal Institute of Technology (ETH-Zurich), CH-8093 Zurich, Switzerland; email: kovac@phys.ethz.ch}     
\altaffiltext{3}{Max-Planck-Institut  f\"ur  Astronomie,  K\"onigstuhl
17,   D-69117   Heidelberg,   Germany} 
\altaffiltext{4}{Space    Telescope    Science
Institute,   3700    San   Martin   Drive,    Baltimore,   MD   21218}
\altaffiltext{5}{Present  address:  Arizona  State University,  Tempe,
Arizona 85287; email: Sangeeta.Malhotra@asu.edu, James.Rhoads@asu.edu}
  \altaffiltext{6}{Center   for
Astrophysics, University  of Science  and Technology of  China, Hefei,
Anhui 230026, P. R. China}

\keywords{cosmology: observations -- early universe -- galaxies: evolution -- galaxies: high--redshift -- large-scale structure of universe}

\section{Introduction}

Galaxy clustering  provides a  powerful tool for  testing cosmological
models and  galaxy formation models,  through quantitative comparisons
between  predicted  and observed  clustering  statistics.  The  galaxy
two-point  correlation  function is  the  most  widely  used of  these
statistics, thanks  to its straightforward calculation  and its direct
relationship to  the galaxy power  spectrum.  It has  been established
for decades that the two-point correlation function is reasonably well
described  by  a power  law  over a  range  of  distances between  the
observed galaxies (see pioneering works  of Totsuji \& Kihara 1969 and
Peebles 1974).

Large  redshift surveys  of  galaxies, such  as  the two-degree  Field
Galaxy Redshift Survey (${\rm 2dFGRS;}$  Colless et al.  2001) and the
Sloan  Digital  Sky  Survey   (SDSS;  Loveday  2002)  provide  precise
measurements  of the  clustering signal  for redshift  $z  \approx 0$.
Their  size makes  it possible  to investigate  the dependence  of the
clustering signal  on intrinsic galaxy properties,  such as morphology
or  luminosity. Red galaxies  are clustered  more strongly,  and their
power law  is steeper, compared to  the power law  which describes the
clustering properties  of blue galaxies  (e.g.  Norberg et  al.  2002;
Zehavi et al.   2002, 2004). This conclusion is  in agreement with the
results from  surveys at  intermediate redshifts about  the clustering
properties of  galaxies of different color (Le F\'{e}vre  et al.  1996;
Carlberg  et  al.  1997).  These  surveys  also  detect  redshift
evolution  in galaxy  clustering.  Recently,  surveys have  achieved a
sufficient size and uniformity  to detect the small deviations between
real correlation functions and pure power law fits (Zehavi et al 2004;
Zheng 2004).

Identification of  large high-redshift galaxy  samples using multiband
color selection techniques (Meier 1976;  Madau et al. 1996; Steidel et
al.   1996,  1998)  has  opened  the way  for  studies  of  luminosity
functions   and  correlation   functions  in   the   distant  universe
(Giavalisco  et  al.  1998;  Adelberger  et al.  1998;  Adelberger  et
al.  2000;  Ouchi  et al.  2003;  Shimasaku  et  al. 2003;  Hamana  et
al.   2003;  Brown  et   al.  2005;   Allen  et   al.  2005;   Lee  et
al. 2006).  Galaxies selected in these broad  band photometric surveys
are  expected  to  have  broadly  similar  properties  and  lie  in  a
restricted redshift interval ($\Delta z \sim 1$).

\lya\ emission offers an  alternative method for finding high redshift
galaxies.   These are  typically star-forming  galaxies  with smaller
bolometric  luminosities than  the  usual continuum-selected  objects.
These samples do  not appear to contain substantial  numbers of active
galactic nuclei  (Malhotra et al. 2003; Wang et al. 2004; Dawson et al.
2004).

In the modern picture of galaxy formation, based on the Cold Dark
Matter (CDM) model, galaxies form in dark matter halos which evolve in
a hierarchical manner. Here, the clustering pattern of galaxies is
determined by the spatial distribution of dark matter halos and the
manner in which dark matter halos are populated by galaxies (Benson et
al. 2000; Peacock \& Smith 2000; Seljak 2000; Berlind \& Weinberg
2002). Galaxy surveys provide constraints on the galaxy distribution.
The dark matter distribution is estimated using N-body simulations or
an analytical approach, generally based on the Press-Schechter
formalism (Press \& Schechter 1974) and its extensions (Sheth et
al. 2001; Sheth \& Tormen 2002). The statistical relation between
galaxies and the dark matter halos where they are found can be
described empirically using a ``halo occupation function''
(e.g. Moustakas \& Somerville 2002), which describes the probability
of an average number $N$ galaxies being found in a halo as a function
of halo mass.

In this article we describe the clustering properties of galaxies
selected through their \lya\ emission at $z \approx 4.5$. In section~2
we present the data used in this paper and describe the selection of
the \lya\ candidates.  In section~3 we present the correlation
function analysis and results.  We compare these results to the
prediction of CDM theory in section~4.  A discussion and a summary of
our main conclusions are given in section~5.  For all calculations we
adopt a $\Lambda$CDM cosmology with $\Omega_{_M} = 0.3$,
$\Omega_{\Lambda} = 0.7$, $H_0 = 70 \kmsMpc$ and the power-spectrum
normalization $\sigma_8 = 0.9$. We scale our results to $h = H_0 /
(100 \kmsMpc)$.

\section{The LALA sample}
 
The Large Area Lyman Alpha (LALA)  survey started in 1998 as a project
to  identify a large  sample of  Ly$\alpha$-emitting galaxies  at high
redshifts (Rhoads et al 2000).  Over 300 candidates have been
identified so far at $z \approx
4.5$ (Malhotra  \& Rhoads 2002),  with smaller samples at
$z \approx 5.7$ (Rhoads  \& Malhotra 2001; Rhoads et  al. 2003)
and $z \approx 6.5$  (Rhoads et al.  2004).
The search  for \lya\ emitters is realized  through narrowband imaging
using the wide-field Mosaic camera at Kitt Peak National Observatory's
4m Mayall  telescope. Two  fields of  view of  36' $\times$  36' are
observed, covering a total area of 0.72 deg$^2$ . In this article we
discuss the properties  of the \lya\ emitters selected  from Bo\"{o}tes
field, centered at 14h25m57s, +35$^0$32'  (2000.0) at $z \approx 4.5$. 
Full details about  the survey  and data reduction  are given in  Rhoads et
al.  (2000) and  Malhotra \&  Rhoads  (2002). 
Five overlapping narrowband  filters of width FWHM $\approx$  8 nm are
used.   The central wavelengths  are 655.9,  661.1, 665.0,  669.2, and
673.0 nm,  giving a total  redshift coverage $4.37  < z <  4.57$. This
translates into  a surveyed volume  of $7.3 \times 10^5$  comoving Mpc$^3$
per field (Rhoads et al. 2000).

Corresponding  broadband  images  are  obtained  from  the  NOAO  Deep
Wide-Field Survey (Jannuzi  \& Dey 1999) in a  custom B$_w$ filter and
the Johnson-Cousins  R and I  filters.  Candidates are  selected using
the following criteria.  In narrowband  images candidates have to be 5
$\sigma$  detections where  $\sigma$ is the locally estimated  noise. The
flux density in narrowband images  has to exceed that in the broadband
images by  a factor  of two.  This corresponds to  a minimum  equivalent 
width (\eqw) of Ly$\alpha$ of 80\AA\  in the observer frame, which helps 
cut down foreground  emitters.  Additionally, the narrowband flux
density must exceed the broad band flux density at the $4\sigma$ level
or above.  Finally,  candidates that are  detected in
B$_w$  band image at  $\ge 2\sigma$  are rejected,  as such  blue flux
should not be present if the source is really at high redshift.

These selection  criteria were followed by visual  inspection.  In the
overlapping area of  all 5 narrowband filters, we  selected a total of
151 candidate Ly$\alpha$ emitters.   More information about the sample
is summarized  in Table 1, where  we give the number  of candidates as
detected  in each  of the  filters.   Because the  filters overlap  in
wavelength, many objects were selected in more than one filter.  Thus,
the total of the sample  sizes for the five individual filters exceeds
the size of the merged final sample.  \\

\section{Two point correlation function}

\subsection {The $w(\theta)$ estimation}

The angular correlation function $w(\theta)$ is 
defined such that the probability of finding two galaxies in
two infinitesimal solid angle elements of size $\delta \Omega$,
separated by angle $\theta$, is $\left(1+w(\theta)\right) \Sigma^2
\delta \Omega^2$, where $\Sigma$ is the mean surface density of
the population.  Typically, $w(\theta)$ is measured
by  comparing  the observed  number  of  galaxy  pairs at  a  given
separation   $\theta$  to   the   number  of   pairs  of   galaxies
independently and  uniformly distributed over the  same geometry as
the  observed  field.   A   number  of  statistical  estimators  of
$w(\theta)$ have been proposed  (Landy \& Szalay 1993; Peebles
1980; Hamilton 1993).

We  calculate the  angular  correlation function  using the  estimator
$w(\theta)$ proposed by Landy $\&$ Szalay (1993)
\be  w(\theta) = \frac{DD(\Theta)-2\,DR(\Theta)+RR(\Theta)}{RR(\Theta)} 
\ee 
where $DD(\theta)$ is the number of pairs of observed galaxies with
angular separations in the range $(\theta, \theta + \delta \theta)$,
$RR(\theta)$ is the number of random pairs for the same range of
separations and $DR(\theta)$ is the analogous number of
observed-random cross pairs. Each of these parameters: $DD(\theta)$,
$RR(\theta)$ and $DR(\theta)$ is normalized with the total number of
pairs in the observed, random and cross-correlated observed-random
sample respectively.

Due to the small number of galaxies detected in the individual filters
we perform $w(\theta)$ calculations for the total sample consisting of
151  galaxies (numbers  are given  in  Table 1).  We are  not able  to
resolve galaxies which  are separated by less than 1  arcsecond from each
other,  thus  we used  this  value as  the  smallest  distance in  the
calculation of  number of  pairs. The random  sample consists  of 1000
individual catalogs, which have been generated to have the same number
of objects and the same  geometry as the observed field. Formal errors
are  estimated for  every bin  using  the relation  $(1 +  w(\theta))/
\sqrt{DD}$ as an approximation of  the Poisson variance, which is very
good  estimation of  the noise  in the  case of  $w(\theta)$ estimator
(Landy  \& Szalay 1993).  Our data  show a  strong correlation  in the
innermost bins,  but the  estimated $w(\theta)$ value  approaches zero
rapidly at $\theta \ga 40"$.

It  is generally  assumed that  $w(\theta)$ is  well represented  by a
power  law $A_{w}  \theta^{-\beta}$.  From  the  estimated $w(\theta)$
values for  our data set, we  conclude that there are  not enough bins
with significant power  for us to estimate both  the amplitude and the
slope of  the correlation law.  For further  calculations we therefore
adopt  the fiducial slope  $\beta =  0.8$.  This  value is  within the
range  for published  Lyman break  samples (see,  e.g.,  Giavalisco et
al. 1998), and is moreover  consistent with results for a flux limited
sample of over  $10^5$ low redshift galaxies from  the SDSS (Zehavi et
al. 2004).

\begin{figure}
\epsscale{0.99} \plotone{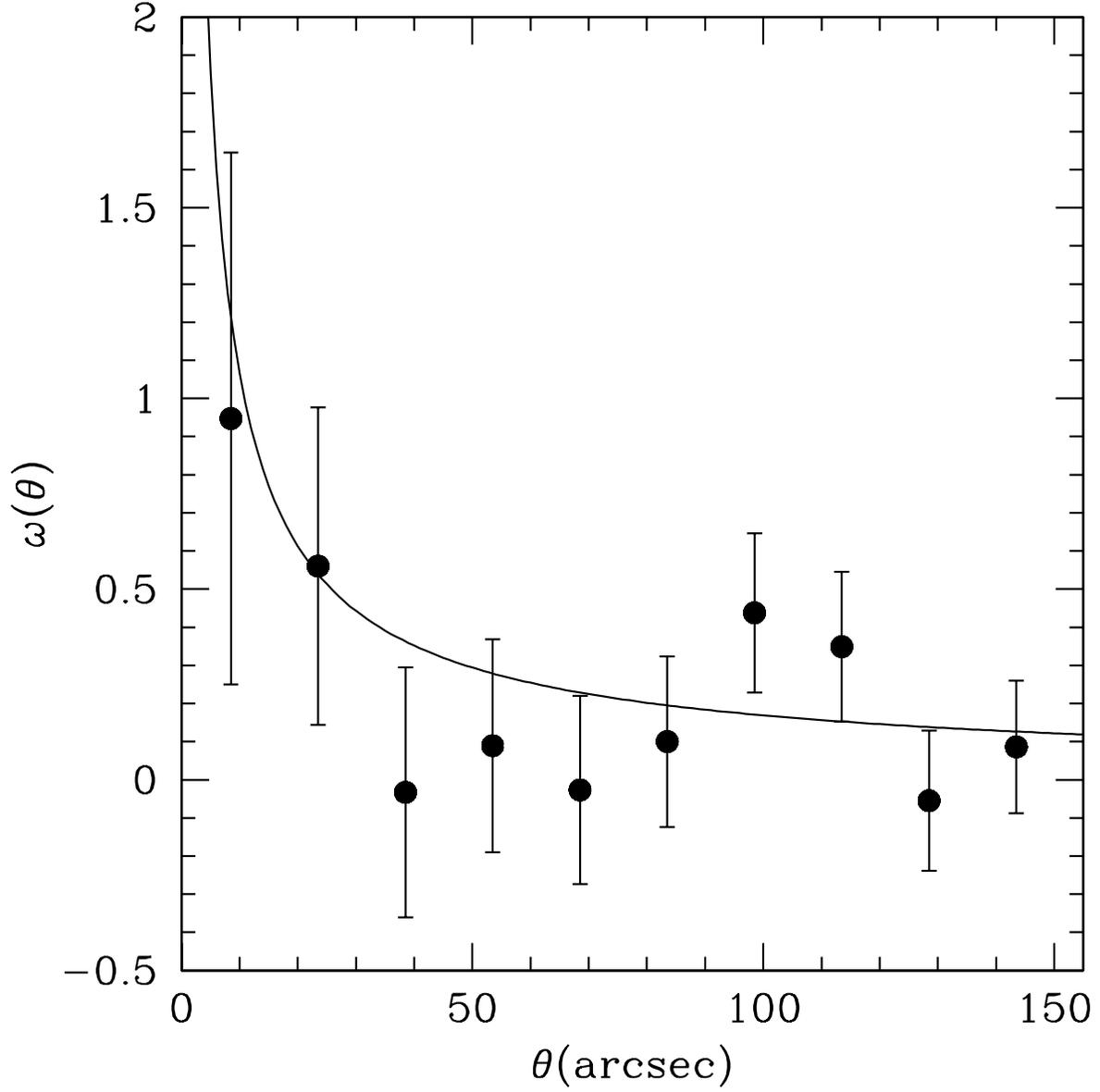}
\caption{The angular correlation function for the sample of 151 \lya\
emitters at  $z \approx 4.5$.  The solid line  is the best-fit  to the
modeled power law $w(\theta) = A_{w} \Theta^{-0.8}$.}
\label{wbootes}
\end{figure}

\begin{figure}
\epsscale{0.99} \plotone{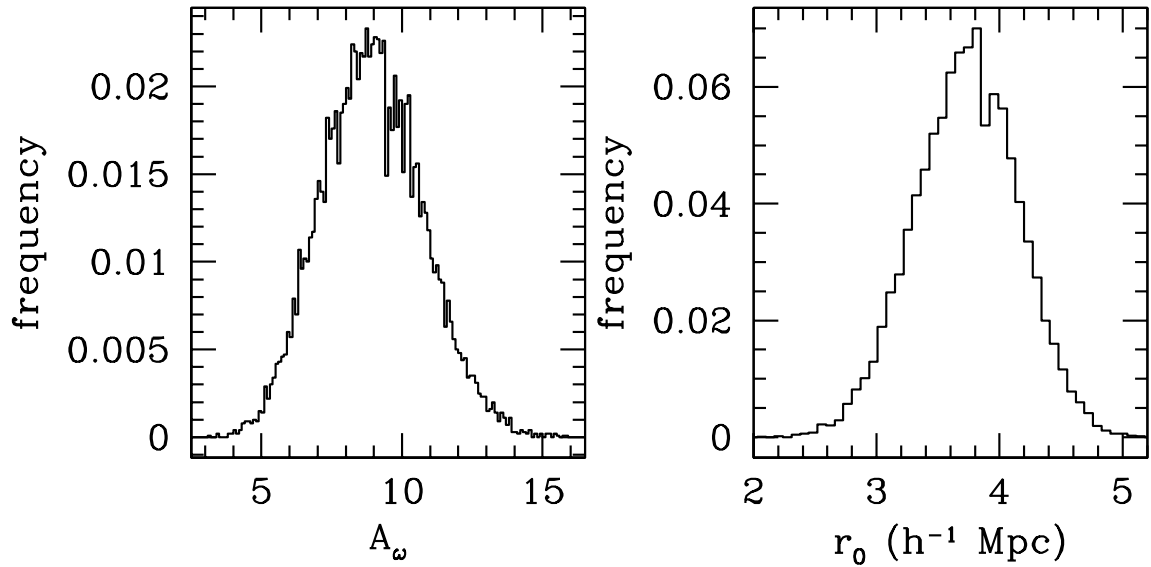}
\caption{Left panel: Histogram  of the best-fit amplitude $A_{w}$
from  the  Monte Carlo  simulation.   Right  panel:  Histogram of  the
spatial correlation length $r_0$,  calculated via Limber equation from
the simulated amplitudes whose distribution is shown in the left
panel.}
\label{hist}
\end{figure}

We use  the $\chi^2$ method to  obtain the amplitude of  the power law
 fitted to  the estimated $w(\theta)$ points, using  the assumed slope
 of $\beta  = 0.8$.  The  best-fit amplitude is $A_{w}  = 6.73\pm1.80$
 for  $\theta$  in  arcseconds (Figure~\ref{wbootes}),  obtained  with
 $\chi^2$=1.90   total  (weighting   the  points   with   the  modeled
 values).  The  confidence  interval  for  the  derived  amplitude  is
 estimated from  the Monte Carlo simulations in  the following manner.
 We create  a set of  10000 random realizations of  $w(\theta)$ values
 modeling them  with a  power law with  the above  estimated amplitude
 $A_{w}$ and  slope $\beta=0.8$ assuming  normal errors (Press  et al.
 1992).   For every realization  of $w(\theta)$  values we  obtain the
 best-fit amplitude using the $\chi^2$ minimization process, fitting a
 power  law with  the fiducial  value  of the  slope $\beta  $.   The
 resulting distribution  of the estimated  amplitudes is given  on the
 left panel of Figure~\ref{hist}.

Estimates of  $w(\theta)$ require  an estimate of  the background
galaxy density.   We base our  density estimate on the  survey itself.
We therefore need to account for uncertainty in the background density
due  to cosmic  variance in  the local  number density  in  our survey
volume. This  bias, known as the ``integral constraint'',  reduces the value
of the  amplitude of the correlation  function by the  amount
(see e.g. Peebles 1980)

\be  C =  \frac{1}{\Omega^2} \int  \int  w(\theta_{12}) d\Omega_1
     d\Omega_2 .  \ee

\noindent
Here $\Omega$ corresponds  to the solid angle of  the survey. The last
integral can be approximated with the expression (Roche et al. 2002)

\be C = \frac{\sum RR \, A_{w} \theta^{-\beta}} {\sum RR} . \ee

\noindent
Summing  over the observed  field we  calculate $C  = 0.00456$. This
value is small and we neglect it in further calculations.

\subsection{The real-space correlation length $r_0$}

In  the  previous  subsection  we  presented  the  measurement  of  the
correlation signal between galaxies projected on the sky. If the
redshift distribution  of the observed  galaxies $N(z)$ is  known, the
spatial  correlation function  can be  obtained from  the  angular correlation function 
using the  inverse Limber transformation (Peebles  1980; Efstathiou et
al. 1991). In the case  of the power law representation of the angular
correlation function, the spatial  correlation function is also in 
power law form and it can be written as

\be 
\xi(r) = (r/r_0)^{-\gamma} .  
\label{xi3d}
\ee

\noindent
The slope $\gamma$ is related to  the slope $\beta$ by $\gamma = \beta
+ 1$.  The amplitudes of  the power law representation of  angular and
spatial correlation functions are related by the equation :
  
\be   A_{w}  =   Cr_0^{\gamma}  \int_0^{\infty}F(z)D_{\theta}^{1-
\gamma}(z)N(z)^2g(z)dz \left [ \int_0^{\infty} N(z)dz \right ] ^{-2} .
\ee

\noindent
Here D$_{\theta}$ is the angular diameter distance,

\be g(z) = \frac{H_0}{c} [(1 + z)^2 (1 + \Omega_{_M} z + \Omega_{\Lambda}
            [(1 + z)^{-2} - 1])^{1/2}] , \ee

\noindent
and C is a numerical factor given by

\be C =\sqrt{\pi} \frac{\Gamma[(\gamma  - 1)/2]}{\Gamma(\gamma / 2)} ,
\ee

\noindent
where  $\Gamma$ stands for  the Gamma  function.  The  function $F(z)$
describes  the redshift  dependence  of  $\xi(r)$, and  we  take $F  =
\hbox{constant}$  given  the  small  redshift  range  covered  in  our
survey.  For the assumed  cosmological model  and the  galaxy redshift
distribution  described  with  a  top-hat  function  in  the  redshift
interval $4.37 < z <  4.57$, we calculate the correlation length $r_0$
of the \lya\ galaxies to be $r_0 = 3.20 \pm 0.42$ $h^{-1}$ Mpc.  The 1
$\sigma$ confidence  interval is  estimated using synthetic  values of
$A_{w}$  created  in  simulations.  The  distribution  of  correlation
lengths shows  smaller scatter than the  corresponding distribution of
amplitudes (Figure~\ref{hist}).

The  observed  clustering signal  may  be  diluted  if our  sample  is
contaminated by  foreground sources. From  the spectroscopic follow-up
of  selected \lya\ emitters  at $z  \approx$ 4.5  the fraction  of the
contaminants  is $f_{cont}  \approx 28  \%  $ (Dawson  et al.   2004).
Presence of  foreground sources can  reduce $A_{\omega}$ by  a maximum
factor  of $(1  -  f_{cont})^2$ assuming  no  correlation between  the
contaminants.  Following this assumption (i.e., no correlation between
the  contaminants)  the  contamination-corrected  spatial  correlation
length for  our sample  is $r_0  = 4.61 \pm  0.60$ $h^{-1}$  Mpc.  The
corrected $r_0$  value corresponds  to the maximum  correlation length
permitted  for  the  sample  studied.  All results  discussed  in  the
following  text  based   on  the  contamination-corrected  correlation
lengths should be therefore understood as the upper limits.

\begin{figure}         
\epsscale{0.5} 
\plotone{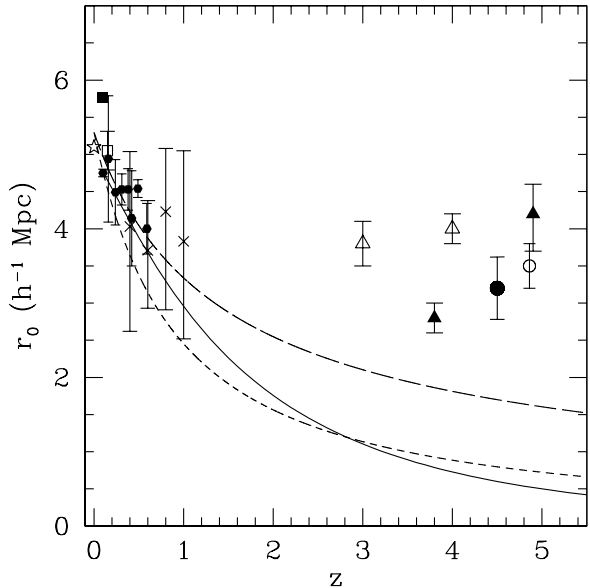}        
\caption{Comparison of  the correlation  length of the  \lya\ emitters
from this  work with correlation  lengths of other  galaxy populations
from the literature. The  filled circle represents our measurement. The
empty circle is the correlation length $r_0$ of \lya\ emitters at $z =
4.86$  from  Ouchi et  al.   (2003).   Triangles indicate  correlation
properties of LBGs.  The open  triangles show measurements for LBGs at
$z = 3$  determined by Adelberger (2000) and at $z  = 4$ determined by
Ouchi et al.   (2004). The last point is for a  sample of the selected
LBGs with $i' < 26.0$. The filled triangles are $r_0$ values by Lee et
al.  (2005) calculated  when both $\beta$ and $A_{w}$  were allowed to
vary. The point at $z = 3.8$ is the $r_0$ value for B-dropouts and the
point at  $z =  4.9$ is the  corresponding value for  V-dropouts, both
with  the  magnitude  limit  $z_{850}   \le  27$  .   The  low- and
intermediate-redshift  measurements of $r_0$'s are represented by empty star
(Loveday et al.  1995;  data from Stromlo-APM redshift survey), filled
square (Zehavi et al.  2002;  SDSS galaxies), empty square (Hawkins et
al.  2003; 2dFGRS galaxies), hexagons (Carlberg et al. 2000; data from
Canadian  Network for  Observational Cosmology  field  galaxy redshift
survey) and crosses (Brunner, Szalay, \& Connoly 2000; data from field
located at 14:20, +52:30, covering approximately 0.054 deg$^{2}$, with
photometrically  measured  redshifts).   The  dashed lines  are  $r_0$
values  as  predicted  by   the  ``$\epsilon$-model''  at  different
redshifts: the short-dashed line  corresponds to parameter $\epsilon =
0.8$  and long-dashed line  corresponds to  parameter $\epsilon  = 0$.
For  comparison the  solid line  shows the  redshift evolution  of the
spatial  correlation length  of dark  matter given  by equation  A1 in
Moustakas  \&   Somerville  (2002).    Having  the  bias   defined  by
equation~\ref{bias} we conclude that high redshift galaxies are biased
more  strong than  the galaxies  from  nearby samples  and samples  at
intermediate redshifts.}
\label{r0compare}
\end{figure}

Figure~\ref{r0compare} shows the observed correlation length $r_0$ (in
comoving units) of  \lya\ galaxies at redshift $z  \approx 4.5$ in our
sample, together  with $r_0$ values  for a number of  surveys covering
the redshift interval $0 < z < 5$. Two points represented with circles
in  Figure~\ref{r0compare} are  measures of  the  correlation strength
from  the  two  samples  of  \lya\ galaxies.  The  correlation  length
estimated  from  our sample  at  $z  \approx  4.5$ (filled  circle  in
Figure~\ref{r0compare}) is in very good agreement with the correlation
length $r_0=3.5 \pm 0.3$ $h^{-1}$ Mpc for the sample of \lya\ galaxies
at $z=4.86$ (empty circle in Figure~\ref{r0compare}) obtained by Ouchi
et al. (2003).

A  discrepancy  arises  when  comparing  the  contamination  corrected
correlation  lengths  from these  two  samples.  In  the following  we
address exactly this  issue in more detail.  Ouchi  et al.  (2003) use
Monte Carlo simulations  to assess the contamination of  the sample by
foreground  sources.   Briefly,  by  generating the  large  number  of
sources created  to correspond  to the detected  sources, distributing
them randomly into  the two real broadband and  one narrowband images,
and consequently  using the  same detecting criteria  as for  the real
sources,  Ouchi  et al.  (2003)  find  that  the maximum  fraction  of
contaminants is  about 40\%.  The contamination  by foreground sources
increases the correlation length up  to the maximum permitted value of
6.2 $\pm$ 0.5  $h^{-1}$ Mpc, quoted in Ouchi et  al. 2003, much larger
than  our  maximum permitted  correlation  length  of  4.61 $\pm$  0.6
$h^{-1}$  Mpc. Even  though the  sample of  \lya\ emitters  studied by
Ouchi et  al. (2003) is peculiar  - galaxies studied  in the discussed
paper belong  to a large-scale  structure of \lya\  emitters discussed
into detail  in Shimasaku et al.  (2003) - we believe  that the reason
for  the discrepancy between  the contamination  corrected correlation
lengths lies in the different methods used to estimate the fraction of
foreground sources.  While our  estimate is based on the spectroscopic
follow-up, the fraction of contaminants derived in Ouchi et al.  (2003)
is  based purely  on the  photometric data.  Shimasaku et  al.  (2003)
discuss  the sample of  \lya\ emitters  at $z  = 4.86$,  extending the
sample  presented  in  Ouchi  et  al.  (2003)  with  additional  \lya\
emitters.  These emitters are  detected in  the field  which partially
overlays  and  partially extends  in  the  direction  of the  observed
overdensity    of    \lya\    emitters    studied    by    Ouchi    et
al. (2003). Shimasaku et al. (2003)  use the same criteria as Ouchi et
al.   (2003)  to  define  the  \lya\  emitters,  except  the  limiting
magnitude of the  \lya\ candidates in the narrowband  images, which is
half  a   magnitude  lower.  Shimasaku  et  al.   (2003)  include  the
spectroscopic  follow-up  to  test  their photometric  selection  (the
spectroscopic  sample contains  5 \lya\  candidates). The  fraction of
foreground contaminants  estimated by  Shimasaku et al.   (2003) using
both the photometric and spectroscopic  data is about 20 \%, two times
lower than  the fraction of  low-z contaminants estimated in  Ouchi et
al. (2003).   Using the updated  fraction of contaminants to  be valid
also  for  the  sample  of   \lya\  emitters  discussed  in  Ouchi  et
al. (2003),  the maximum permitted  correlation length of  that sample
would  be $r_0=4.5  \pm  0.4$ $h^{-1}$  Mpc,  assuming no  correlation
between the contaminants.  This value  is again in very good agreement
with our estimate of the maximum correlation length of $r_0 = 4.61 \pm
0.60$ $h^{-1}$ Mpc  corrected for the dilution of  the sample of \lya\
emitters with the low-z galaxies.

However, one  should bare in  mind that the correlation  properties of
the  sample of  \lya\ emitters  studied  by Shimasaku  et al.   (2003)
differs  from the correlation  properties of  the sample  presented in
Ouchi  et al.   (2003).   The angular  correlation  function of  \lya\
emitters at $z=4.86$  is no longer well described by  the power law of
the angular  distance: it is practically  flat taking values $w  \sim$ 1-2 at
distances $\le$ 8 arcmin, except for the point at 0.5 arcmin (Shimasaku
et al.   2004). The authors claim  that the constant  amplitude of the
angular correlation  function is  largerly implied by  the large-scale
structure and the large void regions in the observed field (see Figure
3 in Shimasaku et al. 2003  or slightly modified Figure 3 in Shimasaku
et al.  2004).  Moreover, Shimasaku  et al.  (2004) searched  the same
field for the  \lya\ emitters at redshift $z=4.79$  (using the imaging
in the additional narrowband filter)  and find only weak clustering of
these \lya\ emitters on any  scale. These results point out that there
exists  a large  cosmic  variance of  clustering  properties of  \lya\
emitters on scales of $\sim$ 35 $h^{-1}$ Mpc (Shimasaku et al. 2004).

The   measured  $r_0$   values   of  \lya\   emitters  (presented   in
Figure~\ref{r0compare})  are  comparable to  the  $r_0$ values  of
LBGs. More  generally, the correlation length of sources  observed at high
redshifts are smaller  for about a third of  the $r_0$ values measured
for the nearby galaxies. When corrected for the contamination by low-z
objects, the maximum permitted correlation  lengths of the samples studied
at  high  redshifts  are  practically  consistent with  the  value  of
the correlation length at zero redshift.

Groth $\&$ Peebles (1977) proposed a theoretical model to describe the
redshift  evolution   of  the   correlation  length,  the   so  called
``$\epsilon$-model''.  In comoving units  this model has  the following
form

\be r_0 (z) =  r_0 (z\hbox{=0}) \times (1 + z)^{-(3 +  \epsilon - \gamma) / \gamma}
~~. \label{epsmodel} \ee

\noindent
For the  fiducial slope of the  correlation power law $\gamma  = 1.8$,
the parameter  $\epsilon  = 0.8$  corresponds  to   the  evolution  of
correlation function  as expected in linear perturbation  theory for a
Universe  with  $\Omega  =  1$.  For $\epsilon  =  -1.2$,  the
clustering pattern is  fixed.  We use normalization $r_0 (z  = 0) = 5.3$ 
$h^{-1} \Mpc$ to calculate  the   modeled  redshift  evolution   of  the
correlation length.  The measurements of the correlation length of the
\lya\  emitters do not  follow  the redshift  evolution of  correlation
length  predicted by  the `$\epsilon$-model'  (short-  and long-dashed
lines in Figure~\ref{r0compare}) . We conclude that there is no value
of $\epsilon$ for which equation~\ref{epsmodel} can fit the observed
correlation  lengths measured for the full range $0 \le z \la 5$.
Similar conclusions have been presented by a number of authors (Giavalisco 
et al. 1998; Connolly et al. 1998; Matarrese et al. 1997;
Moscardini et al. 1998). 

This implies that the population of \lya\ galaxies at $4 \la z \la 5$
is much more strongly biased than the low redshift galaxy samples
shown in Figure~\ref{r0compare}.

Figure~\ref{r0compare} can not be straightforwardly interpreted as the
redshift  evolution   of  the  correlation  length,   given  that  the
correlation length of  the \lya\ emitters (and similarly  of the LBGs)
does  not  necessarily  track   that  of  the  general  population  of
galaxies. Typically,  high redshift  systems have been  selected using
the Lyman-break or  \lya\ techniques, which  are sensitive  to detect
galaxies actively  forming stars. Proper  comparison of the  values of
correlation  lengths  of galaxies  at  low  and  high redshifts  would
require  to select  the local  sample using  the same  criteria  as to
detect high  redshift sources.   For example, Moustakas  \& Somerville
(2002) study three  populations of galaxies  (local giant ellipticals,
extremely  red  objects and  LBGs)  observed  at  the three  different
redshifts (z $\sim$ 0, z $\sim 1.2$ and z $\sim$ 3, respectively) with
clustering lengths  of similar  values.  The masses  of the  host dark
matter haloes,  obtained from the clustering  analysis, of populations
observed  at  different epochs  were  different,  implying that  these
populations of  objects do  not have the  same origin.   Therefore the
values  of the  clustering  strength measured  for  the population  of
galaxies residing  at low and  high redshifts (possibly  corrected for
the contaminants) can  not be used to make  definite conclusions about
the  evolution of  the clustering  properties of  all  galaxies.  More
understanding of the evolution of  galaxies can be gained by comparing
the  clustering properties  of  haloes  which can  host  this type  of
galaxies at a specific epoch.

\section{Comparison with CDM}

Using the correlation length and the comoving number density estimated
from the observations of \lya\ emitters at $z \approx 4.5$, we can
constrain the possible masses of the host dark matter halos of the
observed population. We compute the implied `bias' of the
\lya\ emitters, i.e. how clustered they are relative to the underlying
dark matter in our assumed cosmology. Readers should be cautioned that
there are different definitions of bias used in the literature, and
bias is also a non-trivial function of spatial scale. Quoted numerical
bias values depend on these assumptions. We define the bias as the
square root of the ratio of the galaxy and dark matter real-space
correlation functions:

\be 
b \equiv (\xi_g/\xi_{\rm DM})^{1/2}\, 
\label{bias}
\ee 

\noindent
where we have assumed that both the galaxy and dark matter
correlation functions $\xi$ are represented by a power-law, with slope
$\gamma_{\rm g}=1.8$ for the galaxies and $\gamma_{\rm DM}=1.2$ for
the dark matter (as measured in N-body simulations of Jenkins et al. 1998). 
We compute our
bias values at a comoving spatial scale of 3.6 $h^{-1}$ Mpc, which
corresponds to an angular separation of 100 arcsec at $z=4.5$,
approximately the largest scale where we obtain a robust signal in our
measured correlation function, and is the same scale used in several
other recent analyzes (e.g. Lee et al. 2006).

In order to predict the clustering properties of an observed galaxy
population, we must consider both (a) the expected clustering of the
underlying dark matter halos at a given redshift and in a given
cosmology, and (b) the {\it halo occupation function}, or the number
of objects residing within dark halos of a given mass. This function is
dependent on the survey redshift and sample selection method. The halo
occupation function (or distribution) may be parameterized with
varying levels of complexity.  Here, we use a very simple formulation,
following Wechsler at al. (2001), Bullock et al. (2002), and 
Moustakas \& Somerville (2002). We define $N_g(M)$ to be
the {\it average} number of galaxies found in a halo with mass $M$,
and parameterize this via a three-parameter function:

\be 
N_g(M>\Mmin) = \left(\frac{M}{M_1}\right)^{\alpha} .  
\ee

\noindent
The parameter $M_{min}$ represents the smallest mass of a halo that
can host an observed galaxy ($N_g=0$ for $M<\Mmin$).  The
normalization $M_1$ is the mass of a halo that will host, on average,
one galaxy. The slope $\alpha$ describes the dependence of the number
of galaxies per halo on halo mass. Though extremely simple, this
functional form has been widely used and has been found to be a
reasonably good approximation to the halo occupation function
predicted by semi-analytic models and hydrodynamic simulations
(e.g. Wechsler et al. 2001; White et al. 2001). 

We compute the halo mass function using the analytic expression
provided by Sheth \& Tormen (1999):

\be 
\frac{{\rm d}n_h}{{\rm  d}M} = - \frac{\bar{\rho}}{M} \frac{\rm{d}
\sigma}{\rm{d} M}  \sqrt{\frac{a \nu^2}{c}} \left[ 1  + (a \nu^2)^{-p}
\right] \exp\left[ \frac{-a \nu^2}{2}\right] .  
\ee

\noindent
Here, the parameters $a=0.707$, $p=0.30$ and $c=0.163$ are chosen to match
the halo number density from N-body simulations. The parameter $\nu$ is
defined by $\nu \equiv \delta_c  / \sigma$,
where  $\delta_c \simeq  1.686$  is the  critical
overdensity for the  epoch of collapse and $\sigma$  is the linear rms
variance  of  the power  spectrum  on the  mass  scale  M at  redshift
$z$. Sheth \& Tormen (1999) also give the  halo bias  $b_h$ in  the form

\be  
b_h(M)  =  1  +  \frac{a  \nu^2 -  1}{\delta_c}  +  \frac{2  p  /
\delta_c}{1 + (a \nu^2)^p}.  
\ee

Now, the integral of the halo mass function weighted by the halo
occupation function gives the comoving number density of galaxies:

\be    
n_{g}    =    \int_{\Mmin}^{\infty}   \frac{{\rm    d}n_h}{{\rm
d}M}(M)N_{g}(M) {\rm d}M 
\ee

\noindent
Similarly, the integral of the halo bias as a function of mass weighted by
the occupation function gives the average bias for galaxies:

\be 
b_g  = \frac{1}{n_g} \int_{\Mmin}^{\infty}  \frac{{\rm d}n_h}{{\rm
d}M}(M) b_h(M) N_g(M) {\rm d}M.  
\ee

\begin{figure}
\epsscale{0.99} 
\plotone{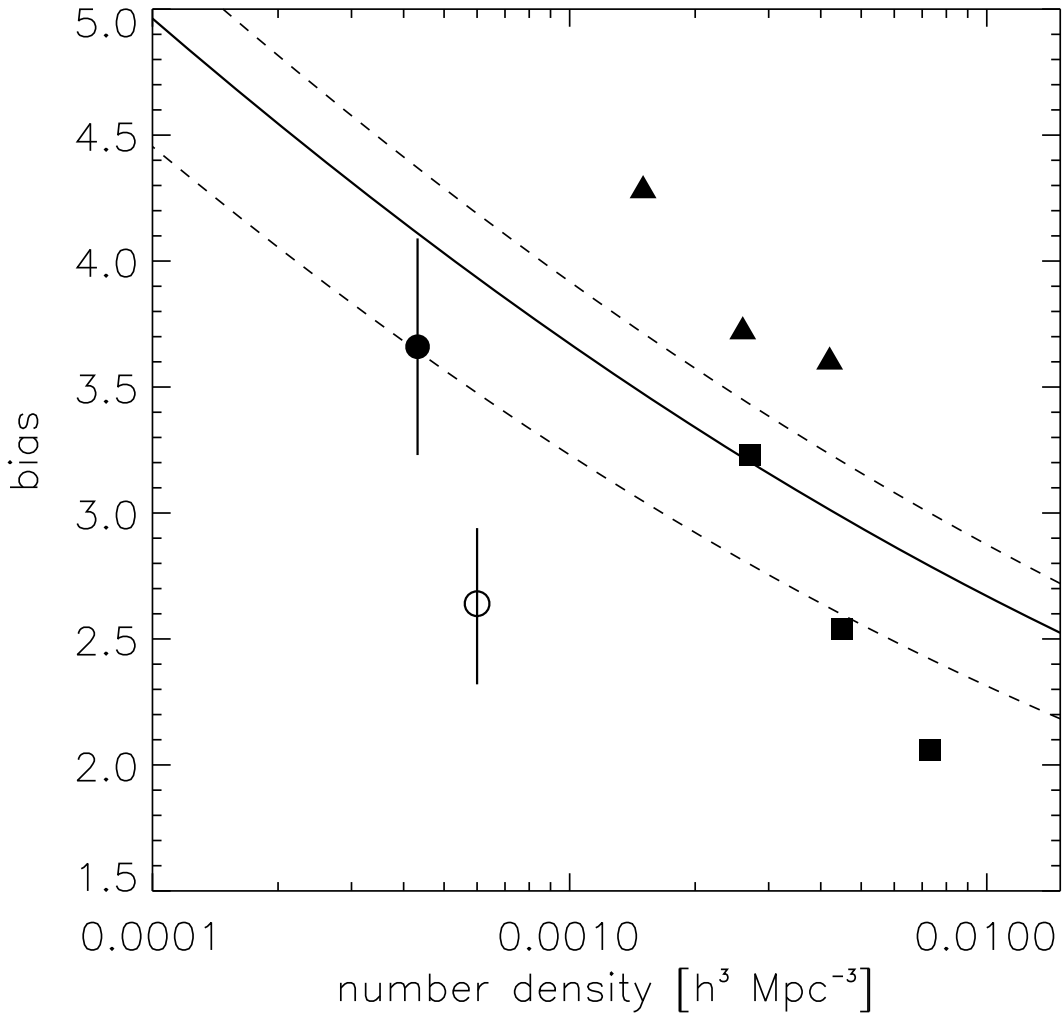}
\caption{Bias vs.\ the comoving number density is shown for our
  observed sample of Lyman-$\alpha$ emitters (open circle: uncorrected; 
  solid circle: corrected for contamination), as well as for dark
  matter halos at $z=4.5$ (solid line). Also shown are number density
  and bias values for Lyman-break galaxies at $z=3.8$ (B-dropouts;
  squares) and $z=4.9$ (V-dropouts; triangles) for three different
  magnitude limits ($z_{850}=26$, 26.5, and 27 from lowest to highest
  number density) from Lee et al. (2006). The dashed lines show the
  relations for dark matter halos at $z=3.8$ (lower curve) and $z=4.9$
  (upper curve) for comparison with the LBGs. The Lyman-$\alpha$
  emitters are less numerous than either dark matter halos or
  LBGs with similar bias values. }
\label{numbias}
\end{figure}

We first  consider the simplest case,  in which each  dark matter halo
above a  minimum mass contains  a single \lya\ emitter  (i.e., $N_g=1$
for $M>\Mmin$).  The comoving number  density and bias values  for the
\lya\ sample,  both uncorrected  and corrected for  contamination, are
shown in Figure~\ref{numbias}, along  with the relation between number
density and  average bias for dark  matter halos as a  function of the
minimum  mass. The  number  density and  bias  values for  Lyman-break
galaxies (LBGs) at $z\sim3.8$ (B-dropouts) and $z\sim4.9$ (V-dropouts)
and for three different observed magnitude limits ($z_{850}=26$, 26.5,
27.0) from the recent study of Lee et al. (2006) are also shown. We 
recalculate the bias values from the Lee et al. (2006) sample using our
definition of bias (Equation ~\ref{bias}); Lee et al. (2006) define the 
bias using the angular correlation function. From Figure~\ref{numbias} 
it is apparent that there is a clear trend for \lya\ emitters to be 
less common than
halos that are as strongly  clustered at their observed redshift. This
may imply that \lya\ is detected  in only a fraction of the halos that
host the objects  producing the emission. It is  also interesting that
the  \lya\ emitters  have  similar  bias values  to  the  LBG
samples  at similar  redshifts,  but again  have  much smaller  number
densities. This suggests a  picture in which the host halos for
these two populations  may have a similar distribution  of masses, but
in which \lya\ emission is seen only a fraction of the time.

We now consider the general  halo occupation function given above, and
invert the equations  for $b_g$ and $n_g$ to  solve for the parameters
$\Mmin$  and  $M_1$. As  noted  by  Bullock et al. (2002),  and  
exploited by  several  recent  studies  such as  Lee  et
al. (2006),  we can  only constrain the  value of the  halo occupation
function slope  $\alpha$ if we  have information on the  clustering of
objects  on rather  small angular  scales. Here  we do  not  have this
information (we have only  one measurement of the correlation function
on scales smaller than 10  arcsec), so our solutions are degenerate in
this parameter.   We give the  values of our obtained  halo occupation
parameters for  three values of  $\alpha$ in Table 2:  $\alpha=0$ (one
galaxy  per  halo),  $\alpha=0.5$,  and $\alpha=0.8$.   We  note  that
Bullock et al. (2002)  found a  best-fit  value of
$\alpha=0.8$ for LBGs at $z\sim3$, while Lee et al.  (2006) found best
fit   values  of   $\alpha=0.65$  and   $\alpha=0.8$   for  $z\sim3.8$
(B-dropout) and $z\sim4.9$ (V-dropout) LBGs, respectively.

We see from Table 2 that the minimum host halo masses range from $\sim
1.6$--$4 \times 10^{10} h^{-1} M_{\odot}$ using the uncorrected values
of number density and bias, and larger values $\sim 1.3$--$2.5 \times
10^{11} h^{-1} M_{\odot}$ for the values obtained when we corrected
for possible contamination of our sample by foreground objects. In
general, $M_1$ is much larger than $M_{\rm min}$, again reflecting
that the \lya\ emitters' number densities are low relative to the
halos that cluster strongly enough to host them.

\section{Discussion and Conclusions}

We have estimated the correlation properties of \lya\ emitters from
the LALA sample at $z \approx 4.5$. From the observed data we measure
the amplitude of the angular two-point correlation function
$A_{\omega} = 6.73\pm1.80$ assuming a fiducial value of the slope of
modeled power law $\beta = 0.8$.  Using the inverse Limber
transformation for the given cosmology and the top-hat redshift
distribution of the analyzed galaxies in the interval $4.37 < z <
4.57$ we calculate the spatial correlation length to be $r_0 = 3.20
\pm 0.42$ $h^{-1}$ Mpc. After correcting for the possible
contamination of the sample by uncorrelated sources (assuming a
contaminant fraction of 28\% based on spectroscopic surveys), we
obtain $r_0 = 4.61 \pm 0.60$ $h^{-1}$ Mpc. This is the maximum
permitted value of the correlation length for our sample.

While large scale structure in the form of voids and filaments is seen
in \lya\ emitters (Campos et al.1999; M{\o}ller $\&$ Fynbo 2001; Ouchi
et  al. 2005,  Venemans et  al.  2002;  Palunas et al.
2000;  Steidel  at al.   2000),  the  measurement  of the  correlation
function is finely balanced between detection (this paper and Ouchi et
al.  2003) and non-detection (Shimasaku et al.  2004). Similar to this
work,  Murayama et  al.  (2007)  measure the  weak  clustering of \lya\
emitters on small  scales (less than 100 arcsec),  which can be well
fitted by a power law. Ouchi  et al. (2004) find correlation at scales
of $\theta  > 50$ $\arcsec$ in a  field where they see  a well defined
clump of \lya\ emitters.  The  distribution of \lya\ emitters from the
survey  of Palunas  et al.   (2004), targeted  on a  known  cluster at
$z=2.38$, show  a weak correlation (significant excess  of close pairs
with separation less than 1 arcmin) and an excess of large voids (size
of 6 - 8 arcmin).  Our detection is at a smaller scale ( $\theta < 50$
$\arcsec$)  in  a  field  with  no noticeable  clumping.   The  spatial
correlation length  we derive agrees  within the 1$\sigma$  error with
the estimate at $z = 4.86$  by Ouchi et al.  (2003), who measured $r_0
= 3.5 \pm 0.3$ $h^{-1}$ Mpc.  On the other hand, the maximum permitted
$r_0$ value  of \lya\  emitters in our  sample is  significantly lower
than the maximum  permitted value estimated by Ouchi  et al. (2003) of
6.2$\pm$ 0.5 $h^{-1}$ Mpc. The  40\% fraction of low-z contaminants in
the  mentioned  work was  derived  using  only  the photometric  data.
Shimasaku  et  al.  (2004)  included  the data  of  the  spectroscopic
follow-up of the enlarged field  observed by Ouchi et al.  (2003), and
derived a  lower fraction of  contaminants of 20\%.  Using  this value
for  the  contamination  by  low-z  galaxies,  the  maximum  permitted
correlation length discussed in Ouchi  et al. (2003) would be $r_0=4.5
\pm   0.4$  $h^{-1}$   Mpc,  assuming   no  correlation   between  the
contaminants.  This fraction of  the low-z contaminants brings our and
Ouchi  et al. (2003)  maximum permitted  correlation length  back into
agreement.

The  $r_0$ values  of \lya\  emitters measured  at high  redshifts are
about 2/3  of the measured  spatial correlation length of  galaxies in
the nearby Universe, or  almost equal when comparing the contamination
corrected correlation lengths of  the discussed \lya.  The high values
of  the  correlation  length  at  high  redshifts,  measured  for  the
specifically selected  samples of galaxies,  which are as high  as the
correlation  length measured  at the  low redshift,  for  more general
populations of galaxies, do not  imply the absence of the evolution in
correlation length.

We compare the measured clustering values with the expected clustering
of dark matter and dark matter halos in the CDM paradigm. We find that
the \lya\ emitters are strongly biased, $b\simeq 2.5$--3.7, relative
to the dark matter on scales of $3.6 h^{-1}$ Mpc at $z=4.5$. These
bias values imply that the \lya\ emitters must reside in halos with
minimum masses of 1.6--$4 \times 10^{10} h^{-1} M_{\odot}$
(uncorrected) or $\sim 1.3$--$2.5 \times 10^{11} h^{-1} M_{\odot}$
using the results after correction for contamination. Interestingly,
 the observed number density of \lya\ emitters is a factor of
$\sim 2$--16 lower than that of dark matter halos that cluster
strongly enough to host them. We further notice that the observed bias
of \lya\ emitters is similar to that of Lyman-break galaxies at $z\sim
3.8$ and $z\sim 4.9$, but again, the number density of the
\lya\ emitters is much lower. This suggests a picture in which the
parent population of \lya\ emitters may occupy dark matter halos with
a similar distribution of masses as those that host LBGs, but are
detectable in \lya\ with a finite duty cycle in the range of 6 to 50\%.

Malhotra \& Rhoads (2002) estimated this duty cycle by combining
stellar population modelling with the extrapolated luminosity function
of LBGs at $z = 4$ (Pozzetti et al.  1998; Steidel et al. 1999).  The
\lya\ emitters were modeled with different stellar population models
to match the observed \eqw\ distribution. To match the number density
of \lya\ emitters, only a small fraction of the inferred number of
faint objects from the LBG luminosity function need to be active in
\lya\ emission. This fraction is derived to be 7.5$\%$ - 15$\%$ ,
depending on the stellar population model, the lower number
corresponding to a zero-metallicity stellar population with an IMF
slope of $\alpha = 2.35$ and whose spectra at the age of 10$^6$ yr is
derived by Tumlinson \& Shull (2000). This is very consistent with the
range of allowed duty cycles inferred from the clustering analysis
presented here. However, the field-to-field variance in the number
density of \lya\ emitters is large, and analysis of more fields is
needed before we can pin this value down further. Measurement of the
correlation of \lya\ emitters on smaller angular scales would allow us
to better constrain the parameters of the halo occupation function, in
particular its mass \mbox{dependence $\alpha$.}

\acknowledgements
This  work made use  of images provided by  the NOAO
Deep Wide-Field Survey  (Jannuzi and Dey 1999), which  is supported by
the National  Optical Astronomy Observatory (NOAO).   NOAO is operated
by AURA, Inc., under a cooperative agreement with the National Science
Foundation.
STScI is operated  by the Association of Universities  for Research in
Astronomy, Inc., under NASA contract NAS5-26555.
We thank Alex S. Szalay, Mauro Giavalisco and Tam\'as  Budav\'ari for useful
discussions, and the  latter also for the help  with the inverse Limber
transformation  calculation. K.K.  would like  to to  thank  STScI for
hospitality during the course of this work.

\label{}
\clearpage

\begin{deluxetable}{ccc}
\tabletypesize{\footnotesize}
\tablewidth{0pt}
\tablecolumns{3}
\tablecaption{Sample statistics}
\tablehead{  
\colhead{Filter} &  
\colhead{Numbers}  & 
\colhead{Surface density (arcsec$^{-2}$)}  } 
\startdata 
All  filters & 151 &  3.51 $\times$ 10$^{-5}$\\ 
H0  & 31  & 7.20  $\times$ 10$^{-6}$ \\  
H4 &  39 &  9.06 $\times$ 10$^{-6}$\\  
H8 & 38  & 8.83  $\times$ 10$^{-6}$\\  
H12 &  66 &  1.53 $\times$ 10$^{-5}$\\ 
H16 & 31 & 7.20 $\times$ 10$^{-6}$ \\ 
\enddata
\end{deluxetable}

\begin{deluxetable}{c c c c c c c c}
\tabletypesize{\footnotesize}
\tablewidth{0pt}
\tablecolumns{8}  
\tablecaption{Correlation statistics parameters}   
\tablehead{
\colhead{}    &  \multicolumn{3}{c}{Measured values} &   \colhead{}   & 
\multicolumn{3}{c}{Halo occupation function parameters}  \\
\cline{2-4} \cline{6-8} \\ 
\colhead{Type of data} &
\colhead{$r_0$ [$h^{-1}$ Mpc]}  &
\colhead{n [$h^3$ Mpc$^{-3}$]} &
\colhead{b}  &
\colhead{} &
\colhead{$\alpha$} &
\colhead{log($M_{\rm min} / h^{-1} M_{\odot}$ )} &
\colhead{log($M_1 / h^{-1} M_{\odot}$) } }
\startdata
&  &  & & & 0 & 10.6351 & --  \\ 
Observed & 3.20  &  6.0 $\times$ 10$^{-4}$ &  2.6 & & 0.5 &  10.44 & 14.76 \\ 
& & & &  &  0.8 & 10.20 & 13.50 \\ 
\hline

Corrected & &  & & & 0 & 11.40  & -- \\
for & 4.61 & 4.3 $\times$ 10$^{-4}$ &  3.7 & & 0.5 & 11.25 & 13.58 \\
contamination & & & & & 0.8 & 11.14 & 12.97 \\

\enddata
\end{deluxetable}

\clearpage

\end{document}